\documentclass[prb,showpacs,floatfix,preprint]{revtex4}
\usepackage{graphicx}
\usepackage{dcolumn}
\usepackage{bm}

\begin{document}

\title{Nonlinear Spin-Charge Dynamics in a Driven Double Quantum Dot}
\author{D.V. Khomitsky$^1$ and E.Ya. Sherman$^2$}
\affiliation{$^1$Department of Physics,
             University of Nizhny Novgorod,
         23 Gagarin Avenue, 603950 Nizhny Novgorod,
         Russian Federation, \\
             $^2$Departamento de Qu\'{i}mica F\'{i}sica,
             Universidad del Pa\'{i}s Vasco-Euskal Herriko Unibertsitatea,
         48080 Bilbao, Spain}

\date{\today}

\begin{abstract}
The coupled  nonlinear coordinate and spin dynamics
of an electron in a double quantum dot with
spin-orbit interaction is
studied semiclassically. The system is
driven by an electric field with the frequency
matching the orbital or the Zeeman resonance in magnetic field.
Calculated evolution of the spin state
is crucially sensitive to the irregularities in the
spatial motion and the geometry of the host nanostructure.
The resulting spin-flip
Rabi frequency has an unusual dependence on the field
amplitude, demonstrating approach to the chaotic spin motion.
In turn, the orbital dynamics depends strongly on the spin evolution due to the
spin-dependent term in the electron velocity.
\end{abstract}

\pacs{72.25.Dc,72.25.Pn,73.63.Kv}

\maketitle

\section{Introduction}

Spins of electrons confined in nanostructures
show a rich coherent and dissipative dynamics (for example, \cite{Valin02,Levitov03,Meza2008})
which should be taken into account for the fundamental understanding
of their properties and possible application in quantum information \cite{Burkard99}
and spin-dependent charge transport. The spin-orbit (SO) coupling, which
usually manifests itself in low-dimensional systems as the linear in the
electron momentum effective field acting on electron spin, is one of the main causes
of this richness. This momentum-dependent coupling offers a new way of manipulating
the electron spin by influencing the electron momentum, that is by an external periodic
electric field. This effect, the electric dipole spin resonance,\cite{Rashba62}
is in the basis of the very promising proposal of  Rashba and Efros \cite{Rashba03}
for spin manipulation of electrons confined by and external potential in lateral quantum dots (QDs)
on the spatial scale of the order between 10 and 100 nm.
The weak coupling of the electric field to the spin
is sufficient to manipulate spins of confined in QDs electrons and holes \cite{Bulaev2007}
and produce inhomogeneous spin patterns in lateral semiconductor structures.\cite{Khomitsky09} The efficiency of this technique  was
demonstrated experimentally with the single-electron QDs.\cite{Nowack07,Pioro08}
Another side of this coupling is the influence of the spin state and dynamics
on the charge motion. This effect, studied experimentally and theoretically
(see, for example \cite{Rokhinson04,Schliemann05, Reynoso08})
is important, especially when it is accumulated at long times
leading to qualitatively new transport phenomena or transitions of the system
to well-separated final states.
The feedback SO coupling effect in the electron weak localization
was considered in \cite{Richter05,Tserkovnyak09}.
Relaxation of spin states due to the coupling to the environment \cite{Fabian2005,Semenov04}
strongly limits the abilities to initialize, keep, and manipulate the desired spin states.

A fast spin manipulation, often on the time scales strongly limited by the spin relaxation times,
requires a relatively large electric field amplitude. This electric field,
in turn, can strongly influence the orbital dynamics
and drive the system far beyond the linear motion in the confining
potential.

A single QD can be well described by a two-dimensional (2D) parabolic or elliptic potential,
however, considerably more complex systems are required for applications, such as, at least,
double quantum dots with a double-minimum potentials.
Taking into account the interest in spin systems in
electric field, the full coupled dynamics of the spin and coordinate degrees of freedom
in a driven QD deserves analysis and understanding of the operation regimes.
One of the central questions arising in such a dynamical
problem for multi-minima potentials is the irregularity and possible transition to
the classical and quantum chaos not only for the
coordinate \cite{Chirikov79,Reichl1984,Lin1992} but also for spin.
The problems which has to be addressed are: how possible irregularities
and chaos in the orbital motion influence the spin dynamics and what effects
in the charge dynamics can be caused by the corresponding irregular spin motion ?
The analysis of these effects aims on understanding the
limits of the spin manipulation in nanostructures and set a link between spintronics
and the nonlinear dynamics and chaos theories.

Most of the studies of spin dynamics in systems not
driven by external field have been performed so far for the two-dimensional lateral QDs.
However, one-dimensional systems, including quantum
wires and wire-based QDs are of interest \cite{Sanchez06,Zulicke08,Ulloa2006},
promising for the applications and can be treated by a thorough analysis of
various double-QD problems.\cite{Ulloa2006} In this paper we concentrate on these systems, which demonstrate
all important for the understanding features of the coupled spin-charge dynamics in driven systems.

\section{Semiclassical model for coupled dynamics}

As a model we consider a
particle with mass $m$ in a quartic
potential
\begin{equation}
U(x)=U_{0}\left(-2\left(\frac{x}{d}\right)^{2}+\left(\frac{x}{d}\right)^{4}\right),
\end{equation}
describing a double-minimum
system where the minima are separated by $2d$ in space with the
barrier of $U_{0}$ in energy.
Below we refer to this system as to the double QD (DQD)
system with each region $U(x)<0$ considered as a one-dimensional QD.
For this potential the classical \cite{Reichl1984}
and quantum tunneling-determined \cite{Lin1992} dynamics driven by electric field
have been studied.
Here we concentrate on the classical dynamics by
considering mainly a wide structure with $d=100 \sqrt{2}$ nm and $U_{0}=25$ meV.
This high energy scale is on the order of 300 K, and, therefore, at temperatures
below 100 K the orbital dynamics is only weakly sensitive to the temperature. For this
reason, the  thermal effects on the  electron momentum and kinetic energy, including the spread
and the activated over-the-barrier motion, will be neglected.
The frequency of the single QD small-amplitude
oscillations $\omega_{0}=2\sqrt{2}\sqrt{U_{0}/m}/d$ is $4.96\cdot
10^{12}$ s$^{-1}$ at GaAs electron effective mass $m=0.067m_{0}$
($m_{0}$ is the free electron mass),
the semiclassically calculated number of energy levels
in a single QD $N_{E<0}={4\sqrt{2}}{d\sqrt{mU_{0}}/{3\pi}}$
is close to eight, and the interdot semiclassical tunneling
probability $\exp\left(-{8\sqrt{2}}{\sqrt{mU_0}d}/3\right)$ is
vanishingly small, which allows to treat the dynamics
semiclassically (we put $\hbar\equiv 1$).

To characterize the SO interaction, we use the Dresselhaus type of coupling
present in all the zincblend structures in the form:
\begin{equation}
H_{\rm so}=\alpha_{D}\left({\bm{\kappa}}{\bm{\sigma}}\right),
\end{equation}
where $\alpha_{D}$ is the bulk Dresselhaus constant, $\bm{\sigma}$ are the Pauli matrices,
$\kappa_{x}=\widehat{p}_x(\widehat{p}_y^2-\widehat{p}_z^2)$ with other components obtained by the cyclic permutation,
and $p_i$ are the corresponding momentum components: $\widehat{p}_x=-i\partial/\partial x$, etc. In a one-dimensional structure,
extended along the $x-$ axis, where the transverse motion is quantized, the Dresselhaus
term is reduced to the linear in momentum term $\beta\sigma^{x}\widehat{p}_{x}$, where
$\beta=\alpha_{D}\langle \widehat{p}_y^2-\widehat{p}_z^2 \rangle$ with the quantum mechanical expectation value
$\langle \widehat{p}_y^2-\widehat{p}_z^2 \rangle$ for the ground state of the quantized transverse motion.\cite{Rashba84}
The value of $\beta$ depends on the asymmetry of the system cross-section and vanishes
if it has a perfect square or circle shape. In the GaAs-based structures with the transverse
dimensions of few nanometers, $\beta$  is on the order of 0.1-0.5$\times10^{-9}$ eVcm.

In the presence of the $z$-axis oriented magnetic field $B_{z}$
the 1D Hamiltonian acquires the Zeeman term only, and with the
electric field $E(t)$ applied along the $x-$axis it has the form
\begin{equation}
H=\frac{\widehat{p}_{x}^{2}}{2m}+U(x)-eE(t)x+\beta \sigma ^{x}\widehat{p}_{x}+\frac{g}{2}\mu _{B}\sigma
^{z}B_{z}.  \label{ham}
\end{equation}
Following the approach of Ref.\onlinecite{Rashba03}, the direct Zeeman coupling of the magnetic
field of the electromagnetic wave to the electron spin was neglected in Eq.(\ref{ham}).
The evolution of mean values for variable $X$ is governed by the
equation of motion ${\dot{X}}=i[H,X]$ where $[,]$
stands for the commutator. By applying
the commutation rules for $x$, $\widehat{p}_{x}$ and for the Pauli matrices one
obtains the following coupled equations for the coordinate and spin dynamics:
\begin{eqnarray}
&&{\dot x}={p_x}/{m}+\beta \sigma^x, \nonumber \\
&&{\dot p_x}=-{\partial U}/{\partial x}+eE(t)-p_x/{\tau_p}, \nonumber \\
&&\dot{\bm{\sigma}}=
\left[\left(\omega_{\rm so}\mathbf{x}+{\rm sgn}(g)\omega_{L}\mathbf{z}\right)\times{\bm{\sigma}}\right],
\label{eqmot}
\end{eqnarray}
where $\omega_{\rm so}=2\beta p_x$, $\omega_{L}=|g|\mu_{B}B_{z}$
is the Larmor frequency, and $\mathbf{x},\mathbf{z}$ are the unit vectors, and upper dot
stands for the time derivative.
Since in the chosen range of parameters, the orbital $(x,p_x)$ dynamic
can be treated classically, here the momentum and coordinate are considered as the quantities governed
by the classical equations of motion.
We include here the momentum relaxation with $\tau_p=5$
ps being a typical relaxation time for the conduction
electron. Since $\omega_0\tau_p\gg 1$, this relaxation
influences the dynamics weakly. Also, we neglect the spin relaxation which
occurs on the time scales \cite{Khaetskii00} much longer than the maximum times
considered in this paper. Hence, the spin states studied here can be considered as
long-standing spin states which properties are of particular importance for spintronics.
The system (\ref{eqmot}) should be accompanied by the initial conditions for the position $x(0)$,
momentum $p_x(0)$, and spin components $\sigma^i(0)$ which values are determined in real structures
by the method of the spin state preparation before the driving field is applied.
The typical set of the initial parameters which we consider in the paper describes the spin-up
electron located at the bottom of a particular quantum well.

\section{Dynamics in  electric field}

\subsection{Zeeman resonance: $\omega=\omega_L$}

To begin with we consider analytically the resonance with
$E(t)=-E_{0}\sin(\omega_{L}t)$ at small $E_0$. For the moment, we suppose that
$|e|E_0\ll U_0/d$, such that all nonlinear terms in the force acting on the electron
are small compared to the linear ones and the orbital oscillations $x(t)$ are very close to the harmonic.
Near the spin-flip resonance, we can consider the spin dynamics with the
explicitly spin-dependent  part of the Hamiltonian:
\begin{equation}
H_s=\beta \sigma ^{x}\widehat{p}_{x}+\frac{g}{2}\mu _{B}\sigma^{z}B_{z},
\label{ham_s}
\end{equation}
in the two-level model, taking only the resonant
contribution $\exp(-i\omega_L t)$ in $\sin(\omega_{L}t)=\left(\exp(i\omega_{L}t)-\exp(-i\omega_{L}t)\right)/2i$.
In this Rotating Wave Approximation neglecting contributions
of the off-resonance terms,  the Hamiltonian (\ref{ham_s}) can be
written as:
\begin{equation}
H_{s}=-\frac{1}{2}\left(
\begin{array}{ll}
\omega_L & \Omega e^{i\omega_L t} \\
\Omega e^{-i\omega_L t} & -\omega_L
\end{array}
\right),  \label{hamres}
\end{equation}
with the Rabi frequency $\Omega=\beta p_R$ where $p_R=|eE_{0}|\omega_{L}/\omega_{0}^2$ is the
amplitude of the driven momentum. The
solution to the Schr\"oedinger equation with Hamiltonian (\ref{hamres})
yields
\begin{eqnarray}
&&\left\langle \sigma ^{x}\right\rangle =-\sin (\Omega t)\sin \omega_{L}t,
\quad
\left\langle \sigma ^{y}\right\rangle =-\sin (\Omega t)\cos \omega_{L}t, \nonumber\\
&&\left\langle \sigma ^{z}\right\rangle =\cos (\Omega t).
\label{sigmares}
\end{eqnarray}
The spin expectation values (\ref{sigmares}) show the Rabi oscillations in
$\sigma^{z}$ while the amplitude of the $\sigma^{x}$ and $\sigma^{y}$
Larmor precession at $\omega_L$ is modulated with the Rabi frequency.
\begin{figure}[tbp]
\centering
\includegraphics[width=0.25\columnwidth]{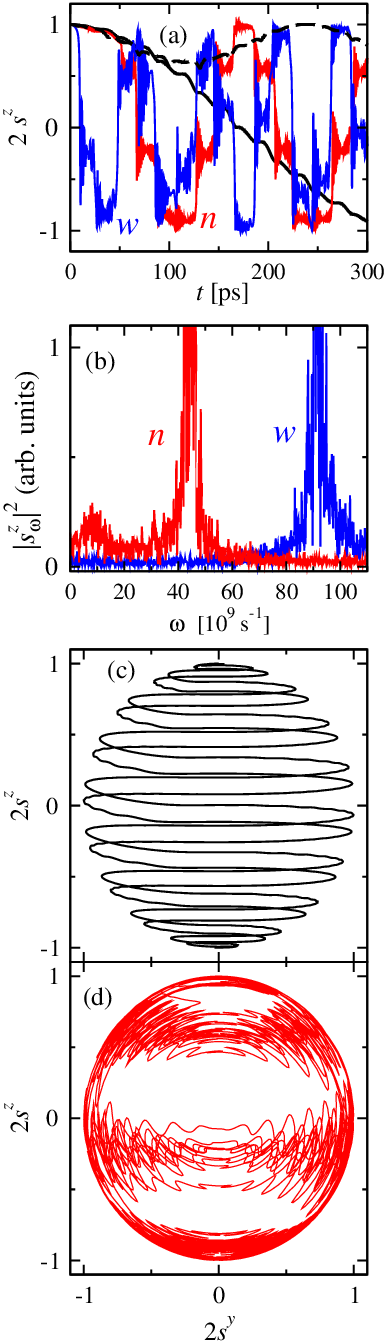}
\vspace{2cm}
\caption{(Color online)
(a) Evolution of $\sigma^z(t)$
at $\omega=\omega_L$ for confined motion in
the left QD  at  $E_0=2.55$ kV/cm (solid black line) and
$E_0=2.6$ kV/cm when the transfer
is induced (line marked $n$, red) for the $d=100\protect\sqrt{2}$ nm structure, and for a
a structure with $d=200\protect\sqrt{2}$ nm with the
transfer at smaller $E_0=1.5$ kV/cm (line marked $w$, blue).
Dashed black line shows $\sigma^z(t)$ at $E_0=2.55$ kV/cm, $d=100\protect\sqrt{2}$ nm,
$\omega=1.1\omega_L$. The full spin flip is achieved only at the exact resonance.
(b) Fourier transform spectra for $E_0=2.6$ kV/cm, $d=100\protect\sqrt{2}$ nm
(line marked $n$, red) and $E_0=1.5$ kV/cm, $d=200\protect\sqrt{2}$ nm (line marked $w$, blue) showing the approach to the irregular spin dynamics.
(c) Phase portrait for $\omega=\omega_L$, $d=100\protect\sqrt{2}$  nm,  $E_0=2.55$ kV/cm, time $t<700$ ps,
(d) same for $E_0=2.6$ kV/cm.
}
\label{Figure1}
\end{figure}

This analytical result can be compared to the numerical solution of
system (\ref{eqmot}) with initial conditions corresponding to
the spin-up electron in the left QD.
Examples of $\sigma^{z}$ dynamics and Fourier power spectra are
shown in Fig.\ref{Figure1} for $g=-0.45$ (we assume the
bulk GaAs value here), $\beta =0.2\cdot 10^{-9}$ eVcm,
and $B_{z}=4$ T corresponding to $\omega_{L}=1.6\cdot 10^{11}$ s$^{-1}$.
The power spectra  $v_{\omega_m}$
were obtained as the Fourier transform of the numerical solution
for the dynamical variables \cite{Yang2007}
$v(t=t_{n})$, $n=1,\ldots ,N$ of (\ref{eqmot})
at the frequencies $\omega_m=2\pi m/t_N$ defined by the overall time
$t_N$ of the observation:
\begin{equation}
v_{\omega_m}=\sum_{n=1}^N v(t_n) e^{-2\pi i (n-1)(m-1)/N}.  \label{fft}
\end{equation}
The time dependence $\sigma^{z}(t)$ is shown in Fig.\ref{Figure1}
(a) for the exact Zeeman resonance $\omega =\omega_{L}$ for three cases
illustrating different regimes of the dynamics, the role of the
structure geometry, and the effect of the interdot transfer for
the spin dynamics and the Rabi frequency.

\begin{figure}[tbp]
\centering
\includegraphics[width=0.25\columnwidth]{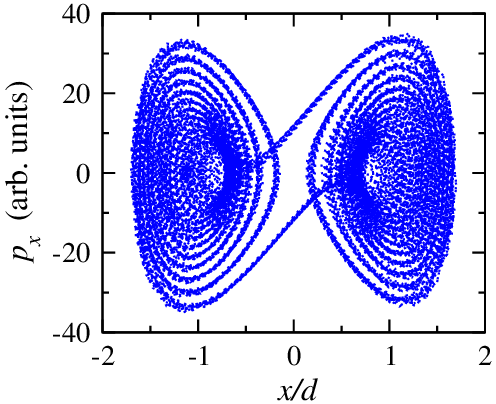}
\vspace{2cm}
\caption{(Color online)
Phase portrait $(x,p_x)$ of the system at $\omega=\omega_L$ and
$E_0=2.6$ kV/cm when the transfer is possible at time $t<700$ ps. The pattern
is dominated by the motion in a single (left or right) QD at the frequencies
on the order of $\omega_0$. The interdot transfer is a relatively rare event
at the frequency $\omega_L$.
}
\label{portrait}
\end{figure}

When the electron is still confined to one dot, the spin motion is regular and
the Rabi frequency is approximately linear in the external field amplitude.
When the amplitude is slightly increased from  $2.55$ kV/cm to $2.6$ kV/cm,
the particle can travel through both QDs causing a strong
change in the Rabi frequency and the entire spin dynamics.
The Rabi frequency in this case is not related directly
to $\omega_{0}$ or the geometrically defined SO frequency $%
\omega _{g}=\beta p_{g}$ ($p_{g}=\sqrt{2mU_{0}}$) since the evolution here is
strongly nonlinear. Above the transfer threshold
the motion is the superposition of two processes shown in Fig.(2): a slow interdot oscillations
with the corresponding frequency $\omega_{L}$ and fast oscillations at the frequencies
on the order of $\omega_{0}\gg\omega_{L}$ in the vicinity
of the minima, not leading to the spin resonance. The increase in the maximum
of the momentum $p_x$ compared the lower-field case, where the transfer
is still prohibited and the motion occurs at the frequencies
determined by $\omega_L$ is on the order of $\omega_{0}/\omega_{L}$.
However, since this very large increase corresponds to the frequencies far
away from the Zeeman resonance, it influences the spin dynamics relatively weakly.
The momentum Fourier component corresponding to the slow motion is proportional
to $\omega_{L}d$ now, causing the abrupt change in the Rabi frequency.
The spin behavior becomes strongly irregular.
In Fig.\ref{Figure1}(b) we present the corresponding Fourier power
spectra in the case of interminimum transfer showing this spin dynamics.
If a broader DQD structure is considered, then a lower driving
field can excite the oscillations with larger electron momentum
due to a more extended and smoother confining potential.
Here the Rabi frequency of $\sigma^{z}$ flip
is increased by a factor of two when comparing the
$d=100\sqrt{2}$ and $d=200\sqrt{2}$ nm structures since the interwell
transfer is characterized by a factor of two greater amplitude of $p_{x}$
at $\omega_L$.
The corresponding phase portraits are presented in Fig.\ref{Figure1}(c) and
Fig.\ref{Figure1}(d) for the regular and irregular dynamics shown in Fig.1.
The plot in Fig.\ref{Figure1}(d) is a signature of approach to
chaotic regime, where the back and forth spin motion on a short time scale
is superimposed by a more regular dynamics leading to the spin-flip on
the longer time scale. It should be noted that the terms "chaotic motion" together
with the "irregular motion" used in the paper refer to the visual shapes of the parameter
dynamics rather than to a strictly defined dynamical chaos. The possible appearance and investigation
of dynamical chaos in our system is outside of the scope of the present paper since our goal here is
limited to observation and description of the transition to irregular dynamics for coupled coordinate
and spin degrees of freedom. Hence, we do not perform a unnecessarily detailed investigation of the dynamical
properties including, for example, the analysis of Lyapunov exponent spectrum.\cite{LiLi}
Still, the analysis presented above and in Fig.\ref{Figure1} certainly lead to a conclusion that
the spin dynamics in our system is strongly affected by the SO coupling and can be tuned by varying
both the geometrical parameters of the nanostructure and the amplitude of external electric field.

\subsection{Orbital resonance: $\omega=\omega_0$}

\begin{figure}[tbp]
\centering
\includegraphics[width=0.25\columnwidth]{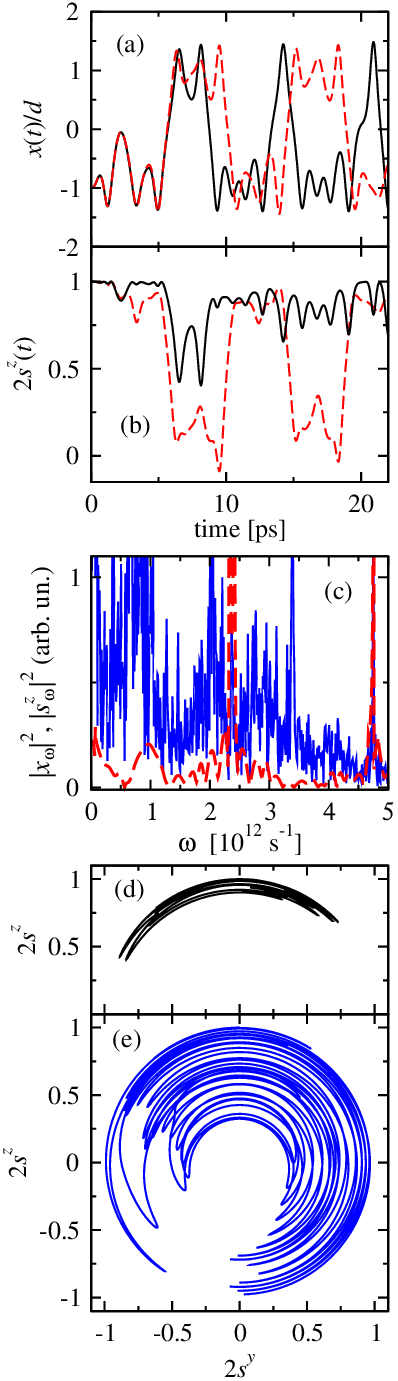}
\vspace{2cm}
\caption{(Color online) Evolution of (a) $x(t)$ and (b) $\sigma^z(t)$ at the primary resonance
$\omega=\omega_0$ and $E_0=1.6$ kV/cm where the particle with $%
g=-0.45$ (solid black line) and $g=4.0$ (dashed red line) is driven through both
QDs. The increase in the $g$-factor leads to much stronger $\sigma^z$ modulation.
(c) The Fourier power spectra corresponding to
$x(t)$ (dashed red line) and $\sigma^z(t)$ (solid blue line) at $g=-0.45$ shown in (a) and (b), respectively.
(d) Phase portrait for $\omega=\omega_0$, $d=100\protect\sqrt{2}$ nm, $E_0=1.6$ kV/cm, time $t<50$ ps
(e) Same for $d=200\protect\sqrt{2}$ nm, $E_0=2.0$ kV/cm, demonstrating the achieved full spin flip.
}
\label{Figure2}
\end{figure}

Another type of dynamics to be considered is the evolution under the
orbital resonance field at the QD oscillation frequency $\omega_0$. When
the electric field amplitude is high enough to overcome the barrier and the momentum
relaxation, the particle can be driven through both QDs which is shown in
Fig.\ref{Figure2}(a) for $E_0=1.6$ kV/cm, although a full flip for $\sigma^z$ is
not achieved even if an artificially large  $g=4.0$ is
considered (dashed lines) as shown in Fig.\ref{Figure2}(b). The
corresponding Fourier power spectra for $g=-0.45$ are shown in Fig.\ref{Figure2}(c).
By comparing Fig.\ref{Figure2} and Fig.\ref{Figure1} it becomes clear that the
expansion in the space motion leads to the enrichment in both coordinate and
spin power spectra due to SO coupling. The dominating part of spectrum is
located between the lowest $\omega_L$ and the highest $\omega_0$ frequencies
of the system which gives rise to various contributions in the whole interval
between them. The phase portrait for $(\sigma^y(t),\sigma^z(t))$ is presented in
Fig.\ref{Figure2}(d) where one can see that the spin dynamics indeed does
not show a full spin flip and is asymmetric with respect to the maximum spin
projections since the potential $U(x)$ is spatially asymmetric with respect to the
single QD minimum.

The character of the spin evolution changes significantly in a broader
DQD structure. Let us consider $d=200\sqrt{2}$ nm which
is twice the initial value and the corresponding frequency $\omega_0$ is now
factor of two lower. By a moderate increase in $E_0$
from $1.6$ to $2.0$ kV/cm one can significantly modify the particle
dynamics and, in particular, achieve a full $\sigma^z$ flip
shown in Fig.\ref{Figure2}(e). The calculations show that
after some period of transient
dynamics a steady $\sigma^z$ oscillations are reached which is further
confirmed by the Fourier power spectrum the dominating
peak at new $\omega_0$. The corresponding phase portrait for $%
(\sigma^y(t),\sigma^z(t))$ with a full spin flip is shown in Fig.\ref{Figure2}%
(e) which comparison to Fig.\ref{Figure1}(d) indicates that the spin evolution
here has a more regular character. This can be
also confirmed by the analysis of $\sigma^z(t)$ dynamics
which shows regular oscillations after the transient period. The results
presented above demonstrate again that the evolution of a spin in the presence of
SO interaction is very sensitive to the spatial motion and geometrical parameters of
the host nanostructure.

\begin{figure}[tbp]
\centering
\includegraphics[width=0.25\columnwidth]{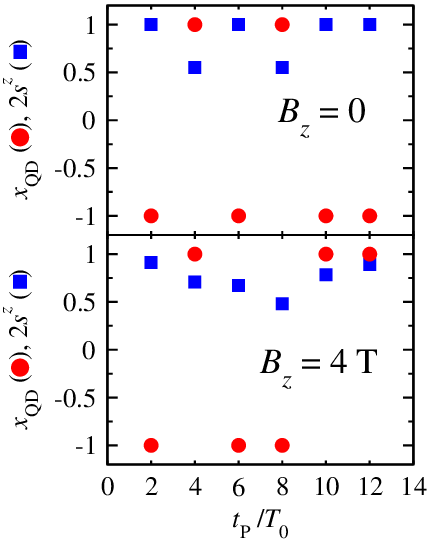}
\caption{(Color online) Asymptotic values $x_{\rm QD}=x(t\rightarrow\infty)/d$ 
(filled circles) and $\sigma^z(t\rightarrow\infty)$ (filled squares) 
after the system has been driven by resonant pulses with $E_0=3$
kV/cm and varying duration for (a) $B_z=0$ and (b) $B_z=4$T. The magnetic
field strongly affects both the spin and the coordinate dynamics demonstrating
the role of SO coupling.}
\label{Figure3}
\end{figure}
To study the long-term effect of the orbital resonance field, a finite length
pulse with
\begin{equation}
E(t)=E_0 \exp \left(-\frac{(t-t_0)^2} {t_{\rm P}^2} \right) \, \sin \omega_{0}t,
\label{epulse}
\end{equation}
will be considered.
We focus on the asymptotic values at $t\gg t_{\rm P}$ for $x(t)$ and $\sigma^{z}(t)$
for the coupled spin-coordinate evolution. Thus, we take $t_{0}=2t_{\rm P}$
and $t_{\rm P}$ varying from $2T_{0}$ to $12T_{0}$, where $T_{0}=2\pi /\omega _{0}$, and the amplitude
$E_{0}=3$ kV/cm. The results are shown in Fig.\ref{Figure3}(a) for the
zero magnetic field and in Fig.\ref{Figure3}(b) for $B_{z}=4$T. It
follows from (\ref{eqmot}) that at zero magnetic field $\sigma ^{x}\equiv 0$
and for the given initial condition $x(0)=0$ the solutions for $x(t)$ and $\sigma^{y}(t),\sigma^{z}(t)$ are related as
\begin{equation}
\left[
\begin{array}{l}
\sigma{^{y}(x)} \\
\sigma{^{z}(x)}
\end{array}
\right] =\left[
\begin{array}{lr}
\cos \left( 2\beta mx\right)  & -\sin \left( 2\beta mx\right)  \\
\sin \left( 2\beta mx\right) & \cos \left( 2\beta mx\right)
\end{array}
\right]
\left[
\begin{array}{l}
\sigma{^{y}(0)} \\
\sigma{^{z}(0)}
\end{array}
\right].
\end{equation}
Therefore, the spin returns to the initial state along with
the coordinate as it can be seen in Fig.\ref{Figure3}(a). When the magnetic
field is applied, the spin-orbit  $\beta\sigma^{x}\widehat{p}_{x}$ and Zeeman $({g}/{2})\mu _{B}\sigma^{z}B_{z}$
terms in Eq.(\ref{ham_s}) do not commute, and such relation is no longer valid. Here,
both the final $x(t)$ and $\sigma^{z}(t)$ values form a different pattern shown in
Fig.\ref{Figure3}(b). We emphasize that not only the spin but the coordinate asymptotic
also becomes magnetic field-dependent due to the SO coupling term $\beta\sigma^x$ in the velocity,
as it can be seen by comparing Fig.\ref{Figure3}(a) and (b).
This result is an important complementary effect of the influence of spin dynamics on
the charge evolution which should be taken into account when both the charge and spin dynamics
in systems with strong spin-orbit coupling are of interest.

\section{Conclusions}

We have studied in the semiclassical approximation
coupled coordinate-spin dynamics of electron in a one-dimensional
double quantum dot with the Dresselhaus type of SO coupling in external magnetic field.
The considered system was driven by a harmonic electric field with the frequency matching
the transition between the orbital levels of the quantum dot or the Zeeman resonance,
or by an intense finite-time orbital resonant field pulse.
The increase in the field amplitude
makes the electron transfer between the potential
minima possible and triggers strongly irregular behavior in both, coupled charge and spin, channels.
These irregularities in the spin motion suggest that in a nonlinear
system evolution of spin coupled to the momentum is very
sensitive to the spatial motion. The spin flips, therefore,
cannot be well described by a single Rabi frequency. In turn, the anomalous spin-dependent
contribution to the electron velocity leads to the position time dependence $x(t)$ strongly
different from that expected for the zero spin-orbit coupling.
The study of the development of these irregularities into the real chaos
in the spin subsystem, involving
the nontrivial Lyapunov spectrum analysis, is an interesting problem for the future
research. The geometry of the nanostructure strongly influences
the spin dynamics in this nonlinear regime. These conclusions emphasize the importance of the possible nonlinear
behavior for the initialization and manipulation of spin states in quantum dots
in the experimentally accessible range of system parameters.

{\it Acknowledgement.} EYS acknowledges support by the Ikerbasque Foundation and the
University of Basque Country (Grant GIU07/40). DVK is grateful to V.A. Burdov and A.I.
Malyshev for helpful discussions and is supported by the RNP Program of Ministry of
Education and Science RF (Grants No. 2.1.1.2686, 2.1.1.3778,
2.2.2.2/4297), by the RFBR (Grant No. 09-02-1241-a), and by the USCRDF
(Grant No. BP4M01)

\end{document}